\documentclass[aps,prl,floatfix,twocolumn,showpacs,tightenlines,groupedaddress,superscriptaddress,amsmath]{revtex4}

\usepackage{graphicx}
\usepackage{graphics}
\usepackage{amssymb}
\usepackage{amsmath}
\usepackage{epsfig}
\usepackage{latexsym}
\usepackage{color}
\usepackage{rotating}
\usepackage{subfigure}
\usepackage{hyperref}
\usepackage{times}

\newcommand{\expect}[1]{\mbox{$\langle #1 \rangle$}}
\newcommand{\ket}[1]{\mbox{$|#1 \rangle$}}
\newcommand{\half}{\mbox{$\frac{1}{2}$}}
\makeatother


\begin{document}

\title{Zeptometer displacement sensing using a superconducting nonlinear interferometer}
\author{Dian Wahyu Utami$^{\text{*}}$, Stojan Rebi\'c, and Jason Twamley}
\affiliation{Centre for Engineered Quantum Systems, Physics
Department, Macquarie University, North Ryde, NSW 2109, Australia}

\begin{abstract}
We propose a design for a superconducting nonlinear interferometer
operating at microwave frequencies which allows the measurement of
the optical nonlinearity $\eta$, with a precision which scales
better than the Heisenberg-like limit as $\delta \eta\sim R^{-3/2}$,
with $R$ the quantification of resources.  By designing the
nonlinear optical element to possess physically moving parts we are
able to use the superconducting nonlinear interferometer to measure
the physical displacement $r$, of the moving parts to a spatial
precision of $\delta (rt) \sim 10^{-21} {\rm m/Hz}$.
\end{abstract}

\maketitle


Recent advances in technology have pushed further the
limits of precision measurement. The standard limit of precision
measurement is given by how the precision scales with the available
resources $R$, such as measurement time or photon number  (see
\cite{Higgins:2007p210} for a general discussion). By utilizing the
quantum properties of a system one can reach a scaling of the
precision $\delta$ that goes beyond the ``Standard Quantum Limit''
(SQL), eg. where $\delta \sim R^{-\alpha}, \alpha > \half$, where
$R$ quantifies the resources required such as measurement time or
photon number. The SQL or ``shot-noise limit'', is achieved when one
attempts to measure an unknown phase using a linear interferometer
with ``classical'' input states. This limit has now been pushed
further to reach a Heisenberg-like limit of $\delta \sim 1/R$ by
various methods. One way to achieve this is by utilizing entangled
states \cite{Giovannetti:2004p11794,Berry:2000p11770}, as input
states into a linear interferometer. Recent experiments approaching,
and also reaching, this Heisenberg-like limit have been performed
with linear interferometers using entangled input states
\cite{Mitchell:2004p11861,Leibfried:2004p11870}, and with feedback
assisted methods using non-entangled   input states
\cite{Berry:2009p10876,Schliesser:2009p9960}. However, achieving a
measurement precision scaling better than $1/R$
 is possible when nonlinear and many-body
interactions are included \cite{Boixo:2008p4824, Woolley:2008p10884,
Choi:2008fe}. Such interactions are typically very difficult to find
naturally and when found are usually very small when compared with
other effects. Recently, a nonlinear interferometer using $\sim 2,000\;^{87}$Rb atoms in a Bose-Einstein condensate held in an optical lattice has experimentally achieved a precision better than the SQL  \cite{Gross:2010p11974}.
In this work we propose a scheme for a superconducting nonlinear microwave interferometer  which incorporates an ultra-large  Kerr
nonlinearity. We show, via homodyne measurements, how to utilize this nonlinear interferometer to
achieve position displacement $r$, sensing with a precision which scales beyond the
Heisenberg-like limit. We show that in-principle
our scheme can achieve a measurement precision which scales as
$\delta r \sim 1/\bar{n}^{3/2}$, where $\bar{n}$ is the average photon number. Further, the degree of precision can be controlled  by adjusting the microwave flux through the interferometer.


The ability to do precise metrology is gaining increasing
importance. Quantum assisted metrology allows one to probe, for
instance, motional displacements of nanoscopic cantilevers
\cite{Teufel:2009p116, Rocheleau:2010hc}, optical phase shifts
\cite{Higgins:2007p210}, sensitive force sensing
\cite{Regal:2008p325}, or magnetic field sensing
\cite{Jones:2009p10885}.  The gravitational wave community has a
special interest in detecting motional displacement and currently
holds the lead in sensitivity for displacement measurement. The
state of the art gravitational wave detection by LIGO (Laser
Interferometer Gravitational-Wave Observatory), which uses 4~km
evacuated beamlines arranged in a high power optical interferometer,
can attain a sensitivity of $\sim10^{-22}\;{\rm m/\sqrt{Hz}}$
\cite{Abadie:2010p11980}. Below we describe how it is possible to
achieve  this level of displacement measurement sensitivity in a
laboratory setup using a superconducting circuit.

\begin{figure}
     \centerline{\includegraphics[width=.80\columnwidth]{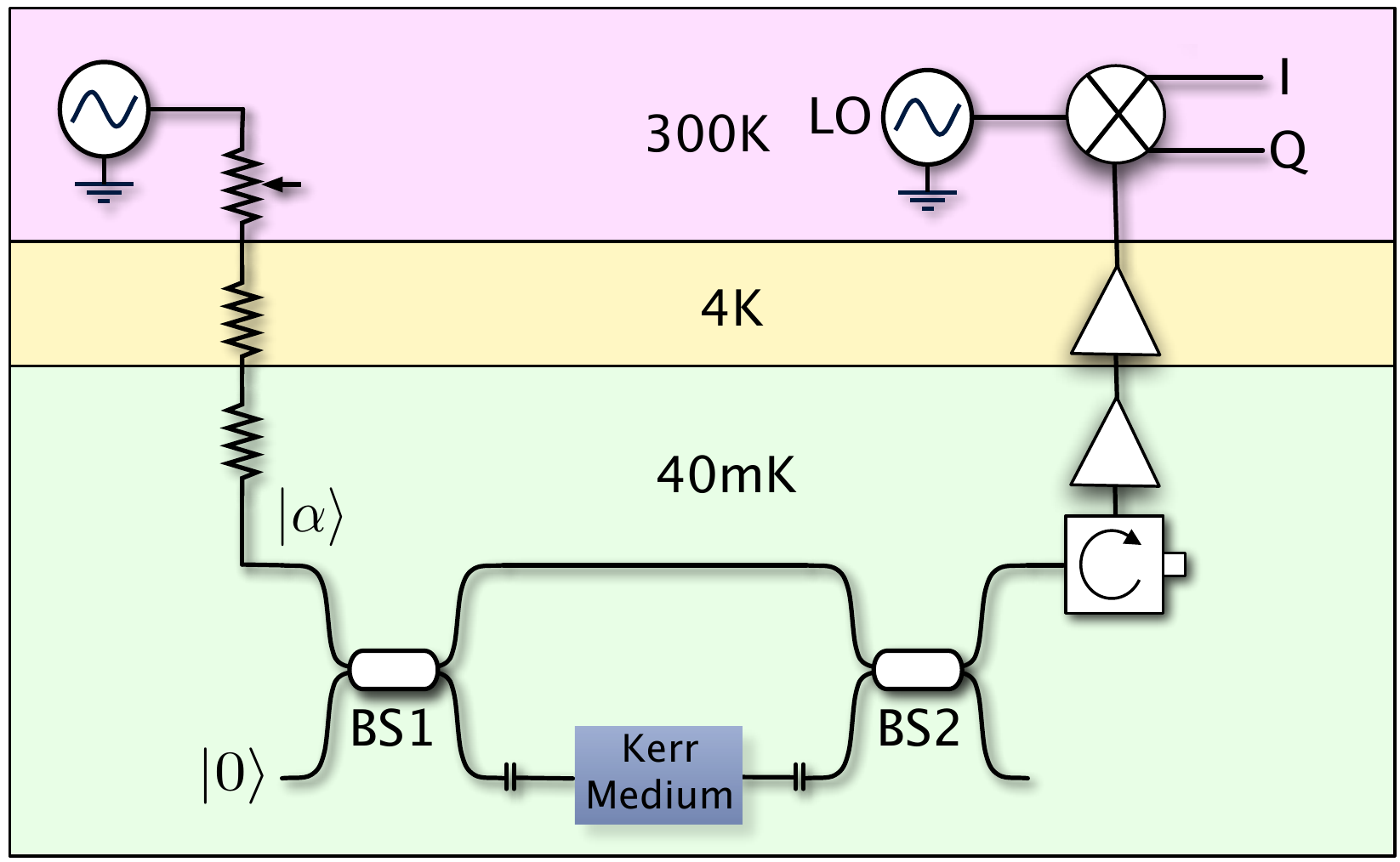}}
  \caption{Schematic circuit diagram of the nonlinear interferometer. One of the interferometer beams passes through a Giant-Kerr medium whose optical nonlinearity depends on a spatial separation within the superconducting circuit and this can be estimated via homodyne detection of an output beam.}
   \label{Fig:Schematic}
\end{figure}

First, as shown in Fig.~\ref{Fig:Schematic}, a coherent microwave
beam is attenuated, passed into a dilution fridge and is split into
two beams by a low-temperature 90${}^\circ$ microwave hybrid coupler
(BS1), while the other input port is left empty (the vacuum). One
output beam propagates through one arm of the interferometer while
the other output passes through a Giant-Kerr nonlinear medium held
within a bad-cavity (defined by the mirror capacitor: $\kappa \gg g,
\gamma$, where $\kappa$
  is the decay rate of EM radiation from the cavity,
  $g$ is the rate of coupling between the cavity and the "atom",
  and $\gamma$ is the decay rate of the "atom").
The two interferometer MW beams are recombined at a second low
temperature hybrid coupler (BS2) and one output from this coupler
 is amplified by a microwave amplifier (e.g. HEMT), and analyzed by homodyne measurements at an
 IQ mixer combined with a Local Oscillator signal.

Our combined device acts like a nonlinear interferometer. Nonlinear
interferometers were first introduced by Kitagawa and Yamamoto
\cite{Kitagawa:1986p5237}, and have been realized optically
\cite{Matsuda:2009p4870},  and more recently via the nonlinear
dynamics of Bose-Einstein condensates \cite{Gross:2010p11974}.

\begin{figure}
     \centerline{\includegraphics[width=.80\columnwidth]{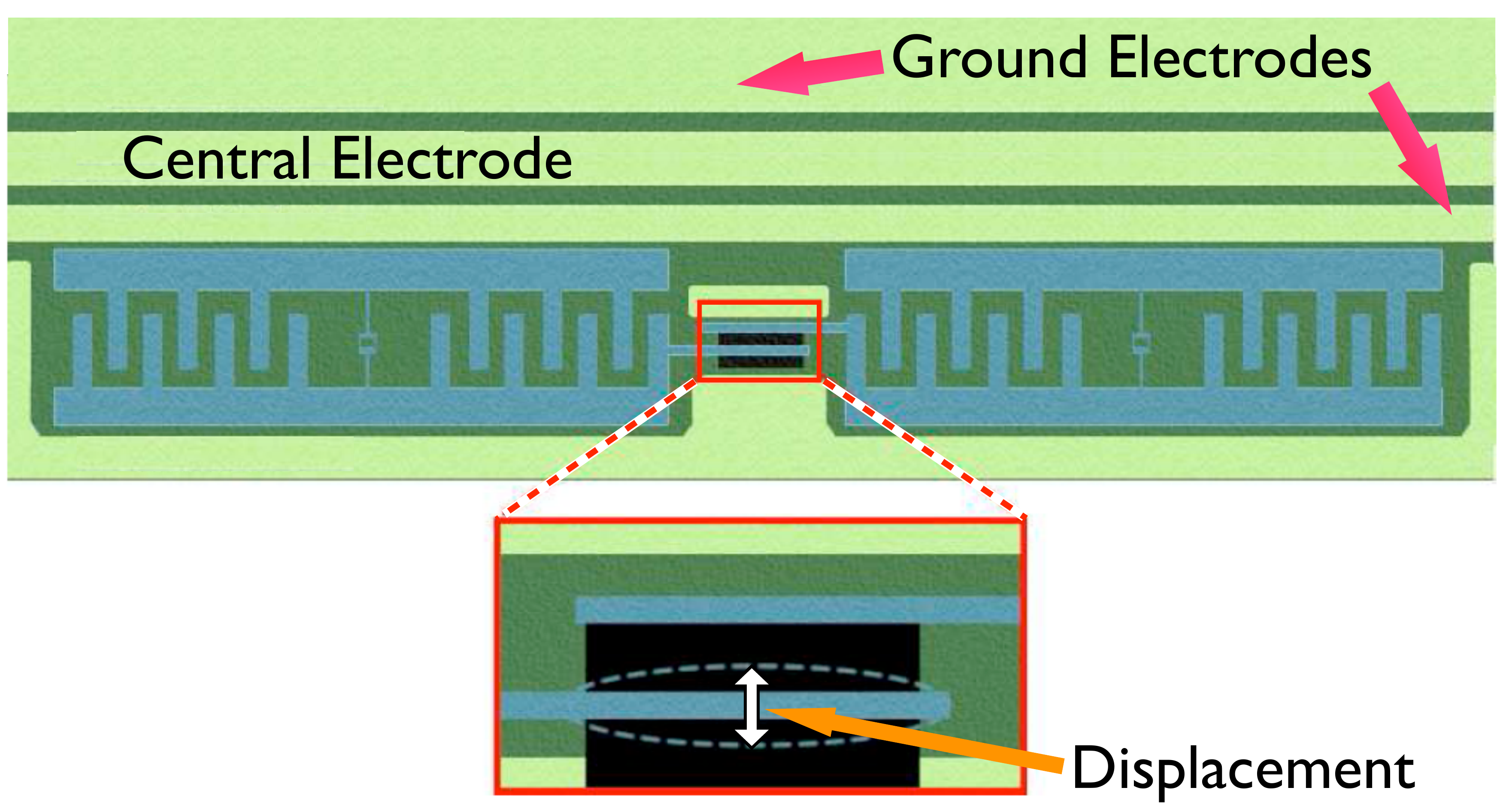}}
  \caption{{Our proposed circuit, consisting two transmon qubits coupled together to a transmission line resonator.}
  Coupling between the transmons is done via the parallel plate capacitors, where the relative distance separation between these two plates can change (as indicated in the inset).}
   \label{Fig:TwoTransmons}
\end{figure}
\begin{figure}
     \centerline{\includegraphics[width=.5\columnwidth]{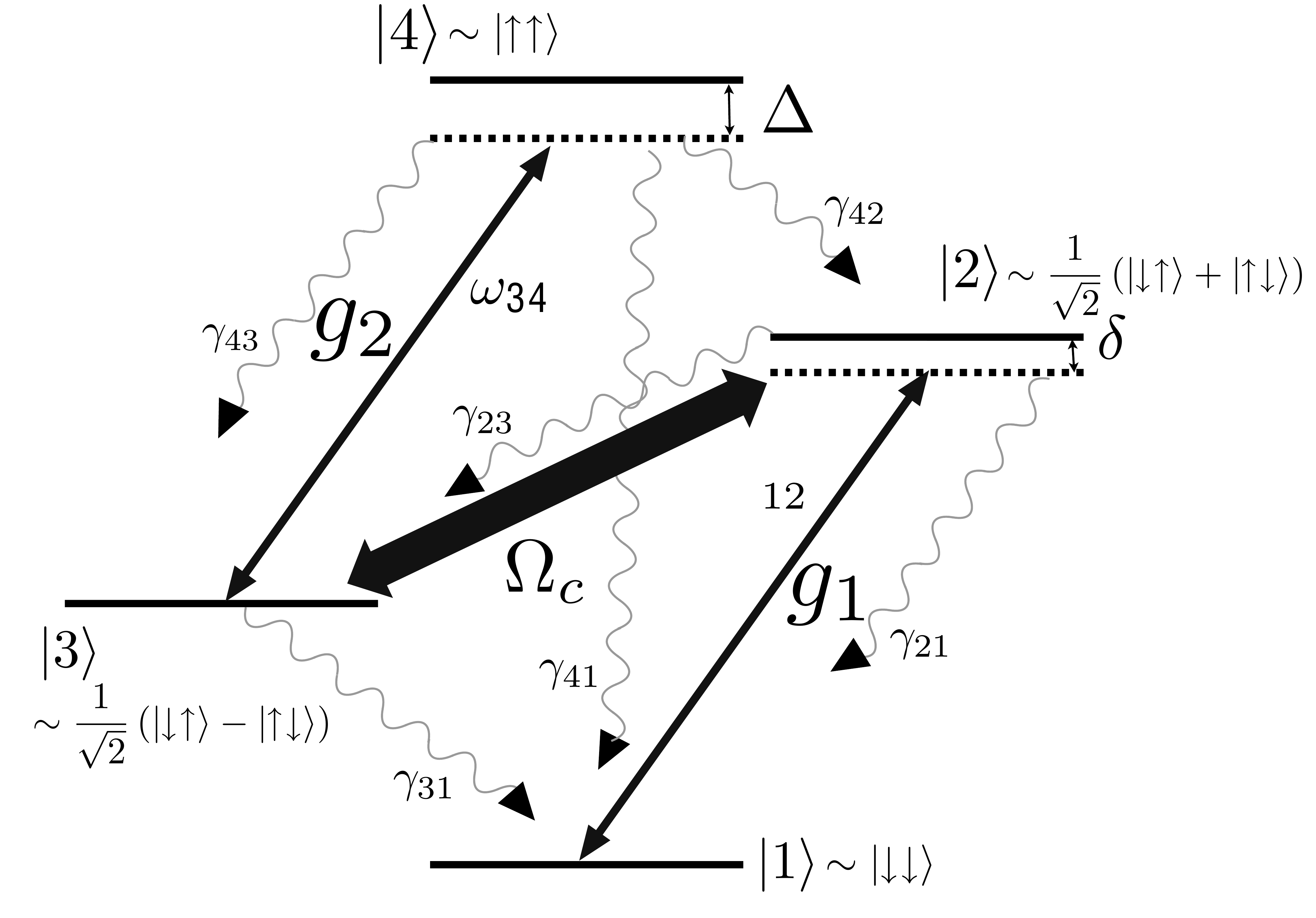}}
  \caption{{4 level scheme in N-Configuration created by coupling the two transmon to a transmission line resonator} \cite{Rebic:2009p10331}}
   \label{Fig:Diagram4levels}
\end{figure}

Our nonlinear Kerr effect is generated through a system that
consists of two superconducting qubits coupled directly to each other and coupled to a
transmission line resonator field mode  creating a 4-level scheme in an
N-configuration \cite{Rebic:2009p10331} (see Fig \ref{Fig:Diagram4levels}).  Although any type of qubit is possible to be employed for this setup, we concentrate on capacitively coupled qubit for ease of adaptation to our application which details would be clear later on. Here we use transmon qubits instead of Cooper-Pair Boxes (CPB) to minimize dephasing noise. The quantised mode of
the transmission line cavity couples to the N-system's transitions $\ket{1} \leftrightarrow
\ket{2}$, and $\ket{3} \leftrightarrow \ket{4}$. In addition, the system is pumped by a classical field
driving the transition $\ket{3} \leftrightarrow \ket{2}$ with a Rabi frequency $\Omega_c$. The pump together with the quantum coupling to the transmission line generates electromagnetically induced transparency for the microwave tone on the $\ket{2}\leftrightarrow\ket{1}$ transition and
consequently eliminates any spontaneous emission from level \ket{2}.
The detuning $\Delta$ on the transition $\ket{3} \leftrightarrow
\ket{4}$ is made larger than the linewidth of the excited state, thus
effecting an ac-Stark shift on the ground state \ket{3} which, in-turn, causes a giant nonlinear optical response for transmission line cavity mode.

It was shown earlier \cite{Rebic:2009p10331}, that the nonlinearity depends on the coupling capacitance through the values of the detunings.
In this work we propose to provide this direct capacitive coupling  via  extended electrodes from each transmon. The simplest possible configuration is
in the form of two capacitor plates as shown in the inset of
Fig.~\ref{Fig:TwoTransmons}. The detunings $\delta$ and $\Delta$, in the
N-system vary with the spatial separation between these two capacitor plates
which in-turn change the Giant-Kerr optical nonlinearity experienced by the microwave tone held in the transmission line cavity. We consider a setup where the capacitor plate separation can change and take the initial setup to be where one plate is fixed in position while the other is movable (see inset of Fig. \ref{Fig:TwoTransmons}).

As shown in \cite{Schmidt:1996p4368}, the N-system, when interacting with the electromagnetic mode trapped in a cavity with loss rate $\kappa$, yields an effective nonlinear Kerr dynamics $H_{eff}\sim \hbar \eta\, a^{\dagger\,2} a^2$, with $\eta$ the Giant-Kerr coefficient. In the limit when  $g_{1,\,2} \ll \Omega_c$, one can find\cite{Rebic:1999p6342}:
\begin{eqnarray}
\eta=\biggl( \frac{g_1}{\Omega_c} \biggr)^2 \biggl( \frac{g_2^2
\Delta}{\gamma_{43}^2+\Delta^2} - \frac{g_1^2
\delta}{(\gamma_{21}+\gamma_{23})^2+(\delta^2)}\biggr)\;\;,
\end{eqnarray}
where we observe that the magnitude of the nonlinearity is independent of
$\kappa$. The ratio $\eta/\kappa$ determines if the system is in the
photon blockade regime ($\eta/\kappa \gg 1$) or not, but it is
important to stress that for large $\kappa$ one can still have large
$\eta$ without a photon blockade. We model the microwave tone traveling wave interacting with the N-system as a driven very bad-cavity containing the N-system and consider the limit $\kappa \rightarrow \infty$. In our simulations we consider a driving tone mode with  $E_p=2\pi\times 5$ MHz, $\Delta=\delta= - 2 \pi \times 60$ MHz,  $g_1=g_2=2\pi\times 100$ MHz, $\Omega_c = 2\pi\times1500$ MHz
and the classical driving field $\Omega_c$ is resonant on the $|2\rangle-|3\rangle$ transition.

We model the input tone to the interferometer as a coherent state $|\alpha\rangle$ which is split by 90${}^\circ$ hybrid coupler (beam splitter) whose action on the incoming field is $ \hat{U}_{\rm BS} = e^{-i\theta (\hat{a}^\dagger \hat{b}+ \hat{a} \hat{b}^\dagger)t}$, where as usual, $\hat{a}$ and $\hat{b}$ describes the two incoming fields to the beamsplitter and $\theta$ describes the phase between the transmitted and reflected field.
We now consider one arm of the interferometer to experience a linear phase shift $\phi$ in addition to the Giant Self-Kerr interaction and this corresponds to the operator $\hat{U}_\eta = e^{-i(\phi \hat{a}^\dagger \hat{a}  + \eta (\hat{a}^\dagger)^2 \hat{a}^2) t}.$
We recombine the two beams using
another hybrid coupler. Taking the input quantum state to the nonlinear interferometer to be $\ket{\psi_{in}}=\ket{\alpha,0}$, we obtain the output state as $\ket{\psi_{out}} = \hat{U}_{\rm BS} \hat{U}_\eta \hat{U}_{\rm BS} \ket{\alpha,0}$.

Via a homodyne detection of one of the output modes we effectively
measure the quadratures: $\hat{X}_{1}=(\hat{a}+\hat{a}^\dagger)/2
,\hat{Y}_{1}=-i (\hat{a}-\hat{a}^\dagger)/2$.

From just one of these of the output quadratures, we can estimate the unknown nonlinear phase with a precision of:
\begin{eqnarray}
\delta(\eta t)=\frac{\Delta X}{|d\expect{\hat{X}}/d(\eta t)|}
\end{eqnarray}
where $\Delta X = \sqrt{\expect{\hat{X}^2}-\expect{\hat{X}}^2}$ is
the variance in the measured quantity $\hat{X}$ quadrature  or
similarly for the $\hat{Y}$ quadrature. Note that our interferometer
scheme uses homodyne detection which is regularly used in
superconducting circuits rather than photon counting, which has only
recently been developed in superconducting systems
\cite{Johnson:2010p8363}.

The quantum limit of measurement using interferometry was first explored
by Caves \cite{Caves:1981p11880}.
We can derive an
analytical expression for the precision of estimating $\eta t$, using either the $\hat{X}$ or $\hat{Y}$
quadratures:
\begin{eqnarray}
\delta(\eta t)(\hat{X})&=&\frac{e^{2 \bar{n} \sin(\eta t)^2}
\sqrt{1+\bar{n}-2 A \cos^2(\phi_1)+B}}{2 \sqrt{2} \bar{n}^{3/2}
|\sin(\phi_1 +2 \eta t)|}\\
\delta(\eta t)(\hat{Y})&=&\frac{e^{2 \bar{n} \sin(\eta t)^2}\text{
}\sqrt{1+\bar{n}-2 A \sin^2(\phi_1)-B}}{2 \sqrt{2}
\bar{n}^{3/2}|\cos(\phi_1 +2 \eta t)|} \label{eq:precXY}
\end{eqnarray}
with $A=e^{-4 \bar{n} \sin (\eta t)^2} \bar{n}$, and $B=e^{\bar{n}
(-1+\cos(4 t \eta ))} \bar{n} \cos(\phi_2)$, where $\phi_1=\phi
t+\bar{n} \sin (2 \eta t)$, $\phi_2=2 (\delta +\eta )t+ \bar{n}
\sin(4 \eta t)$, and $\bar{n}$ is the average photon number
associated with the initial coherent state $\ket{\alpha}$.

In the limit of small time $\bar{n} \eta t \ll 1$, this leads to a
precision $\delta$ which scales as $2 \bar{n}^{-3/2}$. This
precision scales better than the Heisenberg-like scaling of
$1/\bar{n}$.  We now utilize this fact to perform displacement
measurement by arranging the nonlinearity to be highly sensitive to
a spatial displacement.

\begin{figure}
     \centerline{\includegraphics[width=.8\columnwidth]{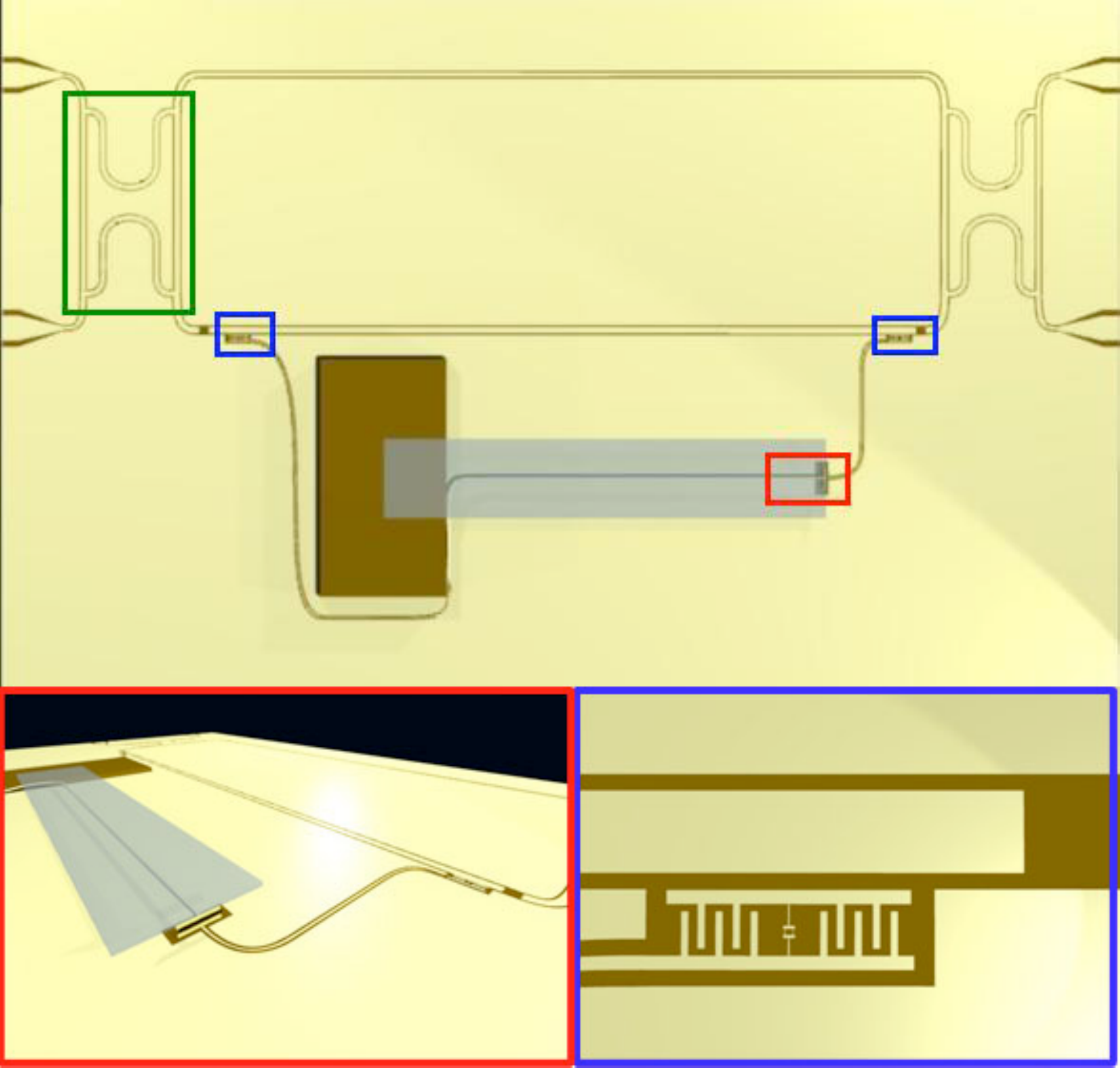}}
  \caption{{ Setup as gravimeter}, showing the signal split into two parts by the 90 degrees hybrid coupler
(green), the two transmon coupled to the transmission line (blue)
and the two capacitive plates, one on a moving cantilever (red).
Lower left figure showing the two capacitive plates, one on the
cantilever, right lower figure showing the transmon in close-up}
   \label{Fig:Allsetup}
\end{figure}

We begin by studying the dependence of the nonlinearity on the
capacitive coupling and particularly on the separation between the
two plates. As an example, we take the two plates as parallelpipeds
of a dimensions  $(w, l, t)= (200,70,0.16) \mu$m. The coupling
capacitance between the two transmons, each with self capacitance of
$100$ fF, now depends on the separation $r$, between these two
capacitor plates.
We consider the Josephson energy of each junction to be
$E_j=2 \pi \times 15$ GHz  \cite{Majer:2007p8156}.  We consider the
transmons to couple to the transmission line cavity with couplings $g_1=g_2 = 2 \pi \times 100$
MHz. From this we can numerically calculate the Giant-Kerr nonlinearity for
various separation distances, as shown in
Fig.~\ref{Fig:Allsetup}a.

 \begin{figure}
     \centerline{\includegraphics[width=.8\columnwidth]{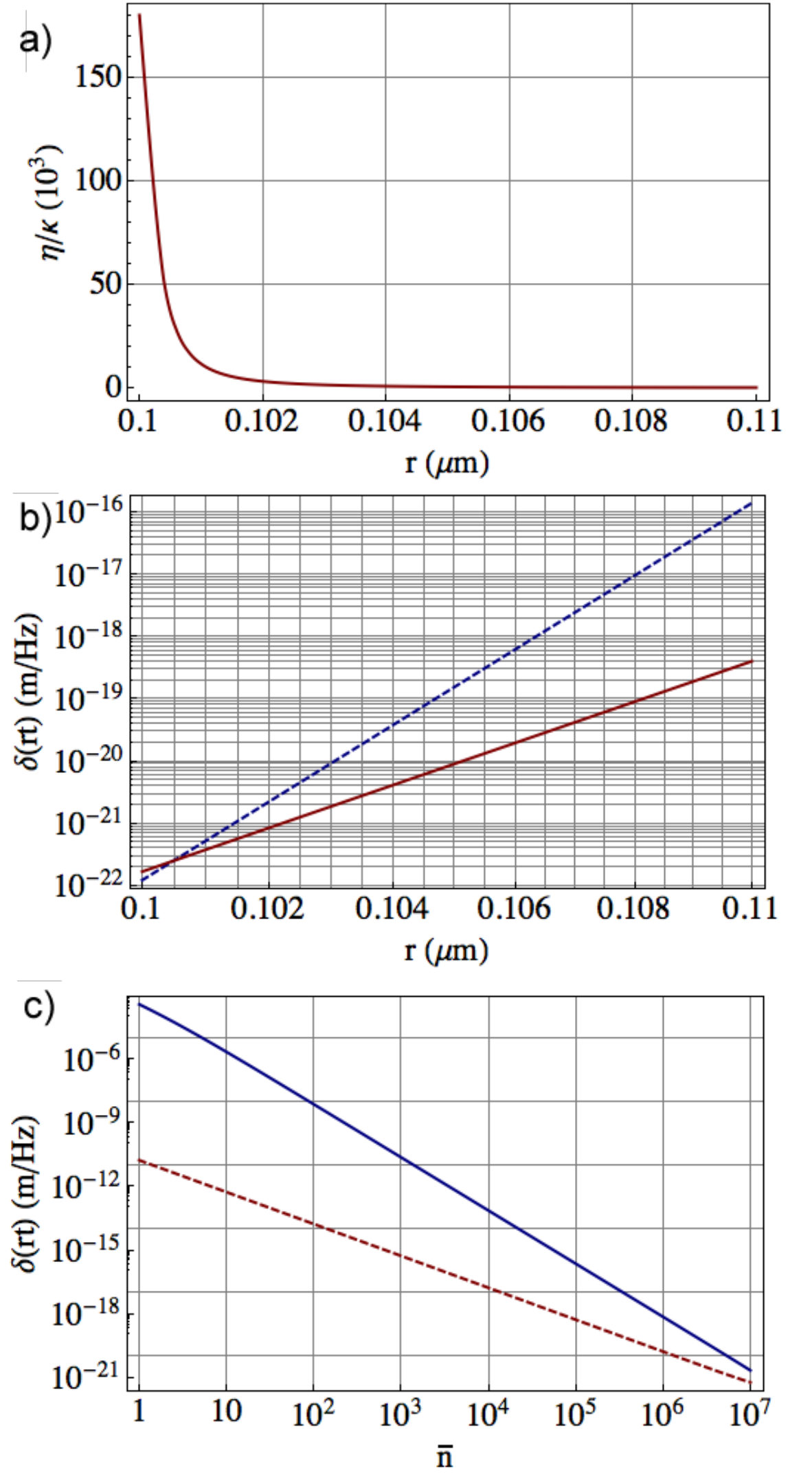}}
  \caption{{(Color online) Dependence of the precision on various variable in the setup.} a). Graph of the Giant-Kerr nonlinearity vs. $r$ when the default plate separation is set at $d=r_0 +r$ where $r_0 = 1.01 \mu$m,
b). Graph of the precision displacement measurement using the
$X$(solid blue) and $Y$(dashed red) quadratures as a function of
$r$.  c). Graph of the precision displacement measurement vs. the
average photon number of the input state for $X$ (solid blue) and
$Y$ (dashed red) quadrature estimation }
   \label{Fig:allgraphs}
\end{figure}

As can be seen from Fig.~\ref{Fig:allgraphs}a,  a nonlinearity of
$\eta/\kappa$ in the order of $10^3-10^4$ can be obtained. Note that
the graph is of ratio of $\eta$ to $\kappa$, thus dependence of the
nonlinearity on the distance can be achieved independently of
$\kappa$.

The precision one achieves when measuring the nonlinear phase via the
quadrature measurement (equation ~(\ref{eq:precXY})),  can be translated into a precision to estimate small changes in the plate separation $r$, as
\begin{eqnarray}
\delta(rt)&=& \frac{\Delta X}{d \expect{\hat{X}}/ d(rt)} = \frac{\Delta X}{\frac{d \langle\hat{X}\rangle}{d(\eta t)} \frac{d(\eta t)}{d(rt)}} = \frac {\delta (\eta t)}{d \eta/dr}
\label{eq:deltar}
\end{eqnarray}

The $r$ dependence of $\delta (rt)$ can be plotted (Fig. \ref{Fig:allgraphs}b), and we note that the precision increases as the separation between the two plates is reduced and has a sharp minimum when the separation is comparable to the thickness of the capacitor plates. In the following we will investigate the precision of
estimating position displacements when the two plates are separated by a
fraction of a micron, a separation distance that is experimentally
feasible.

From equations (\ref{eq:precXY}) and (\ref{eq:deltar}), we find that measurement sensitivity increases
with increasing photon number (Fig.~\ref{Fig:allgraphs}b).  From Fig.~\ref{Fig:allgraphs}c, we can see that we can reach a
precision on the order of $\delta(rt) \approx 10^{-21}$m/Hz for
$\bar{n} \approx 10^{7}$, this is a precision scaling comparable to LIGO. We do not consider when $\bar{n} > 10^7$ as the superconducting waveguides themselves display nonlinear effects. The sensitivity for lower photon number input states are also quite remarkable. Note that we have employed a highly lossy cavity to
mimic the traveling wave interacting with the N-system and we find that the sensitivity can be optimized by
changing the transmon capacitances given a range of displacements.  Although we have not yet treated
noise in this system, we believe that this will not hinder the
workability of our scheme.

Current state of the art on-chip displacement transductors are able
to measure $\delta r \sim 1 \rm{pm} /\sqrt{\rm Hz}$ at room
temperature \cite{Unterreithmeier:2010p11985} while commercial cantilevers are
able to measure $50-200 {\rm fN}/\sqrt{\rm Hz}$ \cite{Olympus:URL}. More recently displacement sensitivity of 250 $\times 10^{-18}$ m/$\sqrt{\rm Hz}$ was obtained using SiN nanomechanical oscillator \cite{Anetsberger:2010p11982}.

By attaching our displacement sensor to a cantilever we can also transform our setup into a mass sensing detector.  We consider a setup where the coupling capacitance is given by two plates being brought together by the motion of a cantilever. If this cantilever is massive it will move as a result of a gravitational force due to another object. One plate is arranged  on the cantilever end and wired to one transmon, whilst the other plate is fixed on the substrate and connected to the
other transmon (see lower left of Fig.~\ref{Fig:Allsetup}). Considering
a rectangular cantilever of $(l,w,t) = (200,70,0.8) \mu$m made of
Silicon Nitride ($\rho = 3184$ kg/m$^3$, $E=250$GPa), one has a spring constant of $k=0.3$ N/m.  This
type of cantilever is commercially available and often used in Atomic
Force Microscopy (AFM) \cite{Olympus:URL, Nanoscience:URL}

An additional mass (for instance an additional cube of
gold of 50$\mu$m side-length) may be attached to the top of the cantilever to increase the
deflection of the cantilever to gravitational forces. With  displacement measurement precision of $10^{-22} $m/Hz, our
system would be able to detect a change in force of $6.6 \times
10^{-17}$N/Hz, equivalent to a force produced by a 1 kg mass at a
distance of 1m away from our cantilever over one second. This is also equivalent to
the ability of equivalently detecting $10^{-9} g$ changes in gravity. The current most precise gravimeter
available is a superconducting gravimeter which is able to
detect changes in gravity with a sensitivity of $10^{-12} g$
\cite{Shiomi:2008p12000} .

We assess the effect of zero point motion on our movable cantilever.
The zero point displacement is given by $x_{\rm zpm}=\sqrt{\hbar/2m\Omega}$, where $\Omega$ the frequency of the cantilever, which, for our setup gives a value of $x_{\rm zpm} \approx 10^{-15}$ m with the additional mass attached to our cantilever.  In using our system directly to sense the displacement of LIGO mirror of $10.7$ kgs oscillating at a frequency of 1 Hz, the $x_{\rm zpm}$ will reduce to $\approx 10^{-18}$ m.  This can be further reduced by one or two order of magnitude when parametric squeezing is applied in one of the quadrature \cite{Blencowe:2000}. Thus our sensor can be used to yield near-Zeptometer/Hz displacement sensing for the LIGO mirrors.

In summary, the scheme presented here allows super-Heisenberg-like scaling in the sensitivity of spatial separation
measurement. We specifically show the implementation of our scheme
in a superconducting circuit for displacement measurement with possible applications to gravitometry.
Our technique, using a nonlinear interferometer, perhaps also can be used to ultra-sensitively measure the changes in the probe height in low temperature scanning probe microscopy.

\begin{acknowledgments}
We acknowledge support from the ARC Centre for Quantum Computer
Technology, the European Union FET Projects QUANTIP \& Q-ESSENCE
and the ARC Discovery Project DP0986932.
\end{acknowledgments}



\bibliographystyle{apsrev}

\end{document}